\documentclass[]{article}
\usepackage{eepic}
\usepackage{epic}
\usepackage{times}

\begin{document}

\title{Alpha-particle condensation in nuclei} 
\author{
{P. Schuck$^1$, H. Horiuchi$^{2}$, G. R\"opke$^{3}$, A. Tohsaki$^{4}$} \\
 $^1$ Institut de Physique Nucl\'eaire, \\ 91406 Orsay C\'edex, France \\  
 $^2$ Department of Physics, Kyoto University, \\ Kyoto 606-8502, Japan \\ 
 $^3$ FB Physik, Universit\"at Rostock \\ -18051 Rostock, Germany \\
 $^4$ Department of Fine Materials Engineering, \\ Shinshu University 
 Ueda 386-8567, Japan \\ }
 
\maketitle

\abstract{A round up of the present status of the conjecture that $n\alpha$ 
nuclei form an $\alpha$-particle condensate in excited states close to the 
$n\alpha$ threshold is given. Experiments which could demonstrate the 
condensate character are proposed. Possible lines of further theoretical 
developments are discussed.}
 
\vspace {1.cm}
   In this article we want to report on the present status of our
recent conjecture [1] that at least light $n\alpha$-nuclei may show around 
the threshold for $n\alpha$ disintegration, bound or resonance states which 
are of the $\alpha$-particle gas type, i.e. they can be characterised by a 
self-bound quite dilute gas of almost unperturbed $\alpha$-particles, all in 
relative s-states with respect to their respective center of mass coordinates 
and thus forming a Bose condensed state. Such a state is quite analogous to 
the recently discovered Bose condensates of bosonic atoms formed in magnetic 
traps. Also there the Bose condensate is characterised by all Bosons occupying 
the lowest s-wave state of the mean field potential. Of course, in the atomic 
case the number of atoms can be enormous whereas in the nuclear case we, so 
far, only consider a handfull of $\alpha$-particles. However, as soon as 
$n\gg1$, one can expect condensation characteristics and therefore 
relatively few $\alpha$'s may be sufficient. The situation is similar to 
nuclear pairing where also a very limited number of Cooper pairs are involved 
and yet most nuclei manifest clear signatures of superfluidity.

  Let us shortly repeat what led us to our conjecture. It is well known that 
the only nucleus which shows a well developed $\alpha$-particle structure in 
its ground state is $^8$Be. This singular feature was probably not given the 
attention it desserves. Other nuclei with three, four, ...,$n\alpha$-particles
collapse in their groud states to a much denser system where the 
$\alpha$-particles strongly overlap and probably loose almost totally their 
identity (see, however discussion below). It seems a very natural idea to 
imagine that, when these $n\alpha$ nuclei are expanded, at some low density 
individual almost unperturbed $\alpha$-particles reappear forming a weakly 
self-bound gas-like state with, at most, pairwise two $\alpha$-particle 
correlations of the $^8$Be-type. Since $\alpha$-particles are Bosons, we 
conjectured that these low density states form a Bose condesate [1]. 
At the moment this conjecture is backed by the following findings. 
Based on the idea of an $\alpha$-particle condensate, we made the following 
very simple variational ansatz for the $n\alpha$ wavefunction: \par

\begin{equation}
< {\bf r}_1\sigma_1\tau_1,...{\bf r}_{4n}\sigma_{4n}\tau_{4n} | \Phi(n\alpha) >
= {\cal A}[e^{-\frac{2}{B^2}({\bf X}_{1}^2+...{\bf X}_{n}^2)}\phi(\alpha_1)...
\phi(\alpha_n)]
\end{equation}

\noindent
where ${\bf X}_i = (1/4)\sum_n{\bf r}_{in}$ is the center of mass coordinate
of the i-th $\alpha$-cluster $\alpha_i$. The internal wavefunction of the
$\alpha$-cluster $\alpha_i$ is
$$\Phi(\alpha_i) = exp[-(1/(8b^2))\sum_{m>n}^4({\bf r}_{im}-{\bf r}_{in})^2]$$
The wavefunction of eq. (1) is totally antisymmetrised by the operator
$\cal A$. It is to be noted that the above wavefunction expresses the state
where $n\alpha$ clusters occupy the same 0s harmonic oscillator orbit
$exp[-(2/B^2){\bf X}^2]$ with $B$ an independent width parameter. For example
if $B$ is of the size of the whole nucleus whereas $b$ remains more or less
at the free $\alpha$-particle value, then the wavefunction (1) describes an
$n\alpha$ condensed state for $n\gg1$. On the contrary, for $B=b$ (1) is a
pure Slater determinant. We took the two parameters $B$ and $b$ of (1) as
Hill-Wheeler coordinates and calculated for spherical geometry the $0^+$
spectrum of $^8$Be, $^{12}$C, $^{16}$O. Of course, the total center of mass
was eliminated. For the effective nucleon-nucleon
force the one of Tohsaki [2] was adopted who adjusted parameters of a sum of
two body and three body Gaussian forces to $\alpha-\alpha$ scattering data
and $\alpha$-particle properties long time ago. The Coulomb force was and is,
of course, also included. The force, therefore, ressembles an effective force
of the Gogny type [3], since the 3-body part gives rise to a density
dependence. Thus our theory does not contain any adjustable parameter.
The solution of the Hill-Wheeler equation yields one $0^+$ state for $^8$Be,
two for $^{12}$C, and three for $^{16}$O. Very close agreement with experiment
has been obtained (within some hundred keV) for the ground state energy of
$^8$Be, for the second $0^+$ in $^{12}$C (with respect to the 3$\alpha$
threshold energy), and for the fifth $0^+$ in $^{16}$O (with respect to the
4$\alpha$ threshold energy). For further details, please consult our papers,
either published [1,4] or on the preprint server [5]. The fact that we get
correctly, without any adjustable parameters, three states close to the
$n\alpha$ desintegration threshold in these nuclei, is very non trivial.
For example, it is well known that it is very difficult to obtain the correct
position of the second $0^+$ state in the $^{12}$C [6]. Only sophisticated
resonating group calculations have achieved this [6]. On the other
hand, most recent AMD calculations, though clearly showing the 3$\alpha$
structure of the second $0^+$ state in $^{12}$C, miss its energy by several MeV
[7]. Concerning the radii, our calculations show that $^{12}$C and $^{16}$O
have, in the threshold states, approximately three times the volume of the one
of their ground states. This confirms the dilute gas structure of these
states. Furthermore also deformed calculations for $^8$Be were performed
and preliminary results show that the wavefunction (1) allows to reproduce
the three lowest members of the rotational spectrum of the $^8$Be very
satisfactority [5]. \par
These successes of our theory make us believe that the $\alpha$-condensed
states are real and we conjecture that they may exist even for quite high
numbers of $\alpha$-particles. For example one may speculate of 10
$\alpha$-particles in $^{40}$Ca. One may also think of a $^{40}$Ca+$^{40}$Ca
head on heavy ion collision. Initially the system gets compressed. If the
energy is just right, the decompression may stall approximataly around the
$\alpha$-condensate density and the whole system may decay into
$\alpha$-particles via the coherent state. Of course, there is temperature
and depending on its value  more or less $\alpha$'s may be broken.
Such scenarios are nice dreams. But how to prove that these experimental
$n\alpha$ threshold states really have Bose condensate character?
The difficulty comes from the fact that the condensation occurs in excited
states of those nuclei. To prove Cooper pair condensation one makes pair
transfer from ground state to ground state and sees that the cross section is
enhanced. But how to make an $\alpha$-particle transfer into an $n\alpha$
threshold state? Nevertheless other methods might exist. Imagine that one
excites the condensate, for example to a $2^+$ state. In $^{12}$C such a state
exists at $\sim$ 3 MeV above threshold. The nucleus may then decay into $n=3$ 
$\alpha$'s which can be detected with a multiparticle detector like INDRA.
Their measured velocity distribution should be proportional to the Fourier
transform of the excited condensate density. The velocity distribution 
will therefore
be rather narrow, if it corresponds to a wide density distribution, as assumed
for the condensate. In heavy ion reactions rotational states are excited.
A rotating condensate should have a strongly reduced moment of inertia with
respect to the rigid body value [8]. Other experiments may be imagined to
put the condensate character of the $n\alpha$ threshold states into evidence.
An indirect indication that the ${0_2}^+$-state in $^{12}$C has extended
character may come from the recent measurement of the spin-orbit splitting of
of the extra neutron in $^{13}$C when added to this state (v. Oertzen et
al., this meeting). Indeed the spin-orbit splitting in $^{13}$C
corresponding to the $0_2^+$-state in $^{12}$C is only half of the one of
that corresponding to the $0_1^+$-state. 
A wider and flatter single particle potential in this excited state with 
respect to the one for ground state could possibly explain this fact.

   Adding neutrons to the $\alpha $-condensate is another very interesting
subject. Adding neutrons to ordinary Cooper-pair condensates quickly destroys
superfluidity due to the blocking effect [8]. This also holds true for
neutron-proton Cooper pairing [9]. However, adding neutrons to a low density
Bose condensed gas of deuterons does not affect the deuteron condensate
because the Pauli principle is not effective [9].
With the $\alpha$-particle condensate we may well have
an analogue to the latter situation of the deuterons. Even on the contrary,
some neutron excess may help to stabilise the condensate (v. Oertzen, this
meeting) making the question of the upper limit of the number of condensed
$\alpha$'s an even more intriguing one. Remember that $^{8}$Be is unbound
whereas $^{9}$Be is bound!

   Let us  now come to the question to what extend even the ground state of
$n\alpha$-nuclei may show superfluid characteristics. The contour plots
$E(B,b) =  <\Phi_{n\alpha}|H|\Phi_{n\alpha}>/<\Phi_{n\alpha}|\Phi_{n\alpha}>$
show [1] that the absolute minima of the energy surfaces have an energy gain
of several Mev with respect to the HF-limit $B=b$. However, one must be careful
with the conclusion about this fact. Let us discuss the issue with the more
familiar situation of standard Cooper pairing. In nuclear physics we are used
to call a nucleus superfluid, when the BCS equations have a non-trivial
solution with a finite gap (contrary to infinite matter where a BCS-solution
always exists for arbitrary weak attractive interaction, in finite systems
the BCS equations not always have a solution because of finite level spacings
and shell fluctuations of the single particle level density). As a rule,
therefore, nuclei at or near shell closure show no pair coherence whereas
open shell nuclei are superfluid. The BCS wavefunction $ |BCS>=exp(P^+)|vac>$,
with $P^+ =\sum_k z_k a_{k}^{+}a_{\bar k}^{+}$ the Cooper pair creation operator,
does not conserve particle number. A better variational wavefunction is
therefore the so-called number projected BCS wavefunction
$|PBCS>=(P^+)^{N/2}|vac>$ where N is, e.g. the number of neutrons. Performing
a minimisation of the energy with respect to the parameters $z_k$ with
$|PBCS>$, we will find that a non-trivial solution always exists, i.e. even
doubly magic nuclei will experience a gain in energy with respect to the
HF-approach. Nontheless we will not consider magic nuclei as superfluid.
This illustrates that $|PBCS>$ theory allows to treat two body correlations
in non superfluid nuclei as well as it contains the superfluid case.
We recognise that our $\alpha$-condensed wavefunction (1) is analogous to
number projected BCS and therefore may describe situations with four particle
correlations as well as situations with $\alpha$-particle condensation.
We believe that (1) largely describes $\alpha$-like correlation in the
ground states of $n\alpha$ nuclei, whereas the threshold states can be
considered as condensates. This statement is corroborated by our recent
investigations of $\alpha$-particle consensation in infinite nuclear matter
[10]. There it turns out that $\alpha$-particle consensation ceases to exist
once the density reaches values where the $\alpha$'s start to overlap
substantially. However, in finite nuclei this needs more carefull
investigation because $\alpha$'s on the surface of, say, $^{40}$Ca in its 
ground state, may still experience low density whereas in the core the 
$\alpha$'s are strongly overlapping. In this respect, an intriguing fact of 
a pronounced even-odd effect in the number of deuterons has recently been 
revealed by K.H. Schmidt et al [11] (see also this meeting). 
Indeed light nuclei with an even number of ``deuterons'' ($n\alpha$ nuclei) 
have systematically a lower binding energy than those with an odd number of 
``deuterons''. 
This rather pronounced ``even-odd'' effect hints to pairing of ``deuterons''.
As a matter of fact sometime ago we have already attempted to describe
$\alpha$-particle condensation within a Boson picture of nucleon  pairs
in applying a bosonic BCS approach to the IBMII model [12].
However, still further investigations must be performed before drawing any
definite conclusions. For example, are $\alpha$-particle transfer reactions
enhanced in $n\alpha$ nuclei? Does the band of ground states energies of
neighbouring $n\alpha$ nuclei show rotational or vibrational character? This
in analogy to pair vibrations and pair rotations, well known from the physics
of ordinary pairing [8]. Another item could be whether the rotational ground
state band of, e.g. $^{24}$Mg shows a rigid or a superfluid moment of inertia.
\par
Let us also make some further remarks about the fact that our wavefunction 
(1) yields an energy minimum several MeV lower than the HF limit [1]. One 
may ask the question, whether present Skyrme- or Gogny-type of HF-calculations 
would also get a lowering of several MeV when $\alpha$-type of correlations 
were included. We think that the answer is 'yes' but then the parameters of 
the forces would have to be readjusted. On the other hand, the force of 
Tohsaki [2] used here has been adjusted with correlations included. It 
therefore is not astonishing that our energy minima lie close to experiment 
and the HF-limit higher.
\par
Further light on these questions could be shed in formulating a theory of 
$\alpha$-particle condensation with a coherent, $\alpha$-particle non 
conserving, state $|\Phi_{\alpha}>$, rather than with our number projected 
state $|\Phi_{n\alpha}>$ of eq. (1). For this, we have to introduce a 
Bogoliubov transformation among Fermion ${\it pairs}$, rather than of single 
Fermions like in the case of Cooper pairing. This can be done in introducing 
quasi-pair operators 
$Q_{\nu}^{+}= \sum_{k,k'}[X_{k,k'}^{\nu} a_{k}^{+} a_{k'}^{+} - Y_{k,k'}^{\nu}
a_{k'}a_k]$ 
and demanding the existence of a coherent state by 
$Q_{\nu}|\Phi_{\alpha}>=0$. Minimising the average energy

\begin{equation}
S_1 = <\Phi_\alpha |[Q_{\nu}[H',Q_{\nu}^+]]|\Phi_{\alpha}>/<\Phi_{\alpha}|
[Q_{\nu},Q_{\nu}^+]|\Phi_{\alpha}> 
\end{equation}

\noindent
with respect to the$ X,Y$ amplitudes leads to the eigenvalue problem

\begin{equation}
 <[\delta Q,[H',Q_{\nu}^+]]> = E<[\delta Q,Q_{\nu}^+]>
\end{equation}

\noindent
with $H'= H-\mu N$. For a two body Hamiltonian the double commutator contains one and two body densities. Since the X,Y-amplitudes form a orthonormal set, the relation $Q^+=Xa^+a^+-Yaa$ can be inverted and then the two body densities can be expressed in terms of X,Y and one body densities. The latter can be calculated from a single particle Green's function consistent with the whole scheme.
This leads to an $\alpha$-particle (four-body) gap equation and an 
$\alpha$-particle order parameter 
$\kappa_{1234}=<\Phi_{\alpha}|a_1a_2a_3a_4|\Phi_{\alpha}>$ in analogy to 
pairing of single Fermions. To good approximation the vacuum is given by

\begin{equation}
|\Phi_{\alpha}> \sim ~ exp[ \sum z_{1234}a_1^+a_2^+a_3^+a_4^+]|vac>
\end{equation}

\noindent
with $z_{1234}=(YX^{-1})_{1234}$.  Partly, this formalism, called Self 
Consistent Quasi-particle RPA (SCQRPA) has been worked out in [13]. In general,
the numerical solution of the corresponding equations is very involved. 
However, approximating the wavefunctions by parametrised Gaussians like in (1),
such calculations may seem feasable. Indeed $z_{1234}$ corresponds to a single 
$\alpha$-particle wavefunction in (1):

\begin{equation}
z \sim \kappa \sim exp[-\frac{2}{B^2} {\bf X}^2 ]\Phi(\alpha)
\end{equation}

\noindent
The problem will be to obtain from this ansatz for $z$ the corresponding
expressions for $X$ and $Y$. This would allow to treat practically any
number of $\alpha$-particles, whereas the present number projected ansatz (1)
is limited to a rather low value of $n$ because the explicit antisymmetrisation 
quickly increases in complexity. Another way to treat large numbers of 
$\alpha$'s is to develope an effective theory, considering the $\alpha$'s as 
elementary particles, i.e. structureless bosons. Equations of the 
Gross-Pitaevskii-type or corresponding Hill-Wheeler equations may then be 
developed. The possibility to treat large numbers of $\alpha$-particles is 
important in order to decide where are the limits of a self-bound gas of 
$\alpha$-particles.

A further speculation related to our studies above is that there may also
exist condensates of heavior clusters, particularly of $^{16}$O nuclei.
For example, $^{48}$Cr might possess an excited $0^+$ state close to the
three $^{16}$O threshold analogous to the $0^+_2$ in $^{12}$C, or $^{64}$Gd
may show a four $^{16}$O gas like state. Probably also partial Bose
condensates exist.
For example exciting $^{32}$S a state with a $^{16}$O - core + 4 loosely
bound $\alpha$-particle may appear.
Such states are being analyzed by M. Brenner et al. [14].
This concludes the round up of the present status of $\alpha$-particle 
condensation in nuclei. We indicated present theoretical results hinting to 
the existence of such states in $n\alpha$ nuclei. We discussed consequences of
such states, open problems, and perspectives.

\section*{Acknowledgements}
We thank Y.Funaki and T.Yamada for fruitful discussions and contributions 
to this work.

\end{document}